\documentclass[12pt]{article}
\usepackage[a4paper, total={7in, 10in}]{geometry}
\usepackage[parfill]{parskip}
\usepackage{physics, tensor, float, subcaption}
\usepackage{graphicx}
\graphicspath{ {Plots/} }
\usepackage{jhep-mod}
\usepackage{bm}
\usepackage{soul}
\usepackage{amssymb,amsmath,amsthm}
\usepackage{mathrsfs}
\usepackage[utf8]{inputenc}
\usepackage{enumerate}
\usepackage{bigints}
\usepackage{xcolor}
\usepackage{appendix}
\usepackage{graphicx}
\usepackage{float}
\usepackage{tikz}
\usepackage{setspace}
\usepackage{cancel}
\usepackage{array}
\usepackage{tabulary}
\usepackage{doi}
\definecolor{purple}{rgb}{1,0,1}
\definecolor{lime}{HTML}{A6CE39} 

\newcommand{\red}[1]{{\color{red} #1}}
\newcommand{\blue}[1]{{\color{blue} #1}}

\definecolor{lime}{HTML}{A6CE39}
\newcommand{\orcidicon}{%
	\begin{tikzpicture}
	\draw[lime, fill=lime] (0,0) 
		circle [radius=0.16] 
		node[white] {{\fontfamily{qag}\selectfont \tiny ID}};
	\draw[white, fill=white] (-0.0625,0.095) 
		circle [radius=0.007];
	\end{tikzpicture}
	\hspace{-5mm}
}
\newcommand\orcidFrancesco{{\href{https://orcid.org/0000-0002-4727-8953}{\orcidicon}}}
\newcommand\orcidStefano{{\href{https://orcid.org/0000-0002-7632-7443}{\orcidicon}}}
\newcommand\orcidMatt{{\href{https://orcid.org/0000-0003-1088-6485}{\orcidicon}}}

\newcommand{\be}{\begin{equation}gin{equation}}
\newcommand{\ee}{\end{equation}}

\def\sign{\mathrm{{sign}}}
\begin{document}
\newcommand{\arXiv}[1]{arXiv:\href{https://arxiv.org/abs/#1}{\color{blue}#1}}

\title{\vspace{-25pt}\huge{
\blue{Fully extremal black holes: \\
a black hole graveyard?}
}}


\author{\Large Francesco Di Filippo\!\orcidFrancesco$^1$,}
\affiliation{$^1$ Institute of Theoretical Physics, Faculty of Mathematics and Physics, \newline
\null\qquad Charles University, V Holešovičkách 2, 180 00 Prague 8, Czech Republic \vspace{3pt}}
\emailAdd{francesco.difilippo@matfyz.cuni.cz}
\author{\Large Stefano Liberati\!\orcidStefano$^{2,3,4}$,\\}
\affiliation{$^2$ SISSA - International School for Advanced Studies, Via Bonomea 265, \newline\vspace{-1pt}
\null\qquad34136 Trieste, Italy \vspace{3pt}}
\affiliation{$^3$
IFPU - Institute for Fundamental Physics of the Universe, Via Beirut 2, \newline
\null\qquad34014 Trieste, Italy \vspace{3pt}}
\affiliation{$^4$ INFN Sezione di Trieste, Via Valerio 2, 34127 Trieste, Italy \vspace{3pt}}
\emailAdd{liberati@sissa.it}
\author{\!\!and \Large Matt Visser\!\orcidMatt$^{5,\dagger}$}
\affiliation{$^5$School of Mathematics and Statistics, Victoria University of Wellington, \\
\null\qquad PO Box 600, Wellington 6140, New Zealand.}
\affiliation{$^\dagger$ Corresponding author.}
\emailAdd{matt.visser@sms.vuw.ac.nz}
\renewcommand{\arXiv}[1]{arXiv:\href{https://arxiv.org/abs/#1}{\color{blue}#1}}
\def\L{{\mathcal{L}}}

\abstract{
While the standard point of view is that the ultimate endpoint of black hole evolution is determined by Hawking evaporation, there is a growing evidence that classical and semi-classical instabilities  affect both black holes with inner horizons as well as their ultra-compact counterparts. In this essay we start from this evidence pointing towards extremal black holes as stable endpoints of the gravitational collapse, and develop a general class of spherical and axisymmetric solutions with multiple extremal horizons. Excluding more exotic possibilities, entailing regular cores supporting wormhole throats, we argue that these configuration could be the asymptotic graveyard, the end-point, of dynamical black hole evolution --- albeit the timescale of such evolution are still unclear and possibly long and compatible with current astrophysical observations.

\bigskip
\noindent
{\sc Date:} Thursday 28 March 2024; Monday 13 May 2024; \LaTeX-ed \today

\bigskip
\noindent{\sc Keywords}: Black holes, endpoint of black hole evolution,  extremality.  \\


\bigskip
\noindent
\leftline{\red{\sf Essay written for the Gravity Research Foundation 2024 Awards for Essays on Gravitation}}

\leftline{\red{\sf Essay awarded the Fifth Prize in  the Gravity Research Foundation 2024 Awards}}

}

\maketitle
\def\tr{{\mathrm{tr}}}
\def\diag{{\mathrm{diag}}}
\def\cof{{\mathrm{cof}}}
\def\pdet{{\mathrm{pdet}}}
\def\QED{ {\hfill$\Box$\hspace{-25pt}  }}
\def\d{{\mathrm{d}}}
\def\sign{\hbox{sign}}

\parindent0pt
\parskip7pt

\clearpage
\null
\vspace{-75pt}
\section{Introduction}

While we have a fair idea about how black holes can be produced, we so far still lack a clear picture about their end. In past decades, the standard lore has been that this end can come about solely as a consequence of Hawking evaporation: a semi-classical, slow, process (the typical timescale is of order $1/(m_P^2 \kappa_{\rm outer}^3) \sim m^3/m_P^2$), which would be irrelevant, at least for eons to come, for any astrophysical black hole.

Traditionally, the possible endpoints of such process have been mooted to be described by the trichotomy of naked-singularity/\-complete-evaporation/\-remnant~\cite{Russo:1992,Strominger:1994, Banks:1994ph, Unruh:2017uaw, Hossenfelder:2009xq,Chen:2014jwq}. The point we want to advance in this essay is that it is increasingly clear that this traditional trichotomy is in need of further refinement. 

It is now quite firmly established that there are for this problem at least two relevant classical instabilities, and a semi-classical one, in addition to that introduced by Hawking radiation. The two classical instabilities are the mass inflation of inner horizons and the light ring instability associated with stable light rings that have to be present in any horizon-less remnant that also supports an (outer) standard unstable light ring~\cite{Cunha:2017qtt, DiFilippo:2024, Cunha:2022gde, Guo:2024cts, Franzin:2023slm}. 

The timescale for the first instability is short, it is by now clarified that it grows exponentially in a timescale of order $1/\kappa_{\rm inner}$ where $\kappa_{\rm inner}$ is the surface gravity of the inner horizon~\cite{Carballo-Rubio:2021bpr,Carballo-Rubio:2022kad, Carballo-Rubio:2023kam, Carballo-Rubio:2023mvr, Carballo-Rubio:2022nuj, DiFilippo:2022qkl, Carballo-Rubio:2018pmi}. For example, in Reissner-Nordstr\"om black hole or Kerr black holes, this timescale is expected to be basically set by the black hole mass, and hence be comparable with the light-crossing time. 

Let us add, that it was recently firmly established that this fast instability is not limited to the usual strictly-defined event and/\-or Cauchy horizons of stationary geometries, but also applies to adiabatically evolving non-extremal apparent/trapping horizons~\cite{Carballo-Rubio:2024, Barcelo:2010pj,Visser:2014zqa}. Hence, it is of physical relevance for any kind of compact trapped region characterized by one or more pairs of inner-outer horizons.

Regarding the light ring instability, the relevant timescale is much less well understood, as very few investigation have been carried out so far, and those only for very specific models~\cite{Cunha:2022gde,Guo:2024cts,Franzin:2023slm}. Nonetheless, being a classical instability, there is no reason to expect an $\hbar$-suppressed timescale, as occurs for Hawking radiation. It could still be long for astrophysical phenomena, but possibly short on cosmological scales.

Coming to the additional semi-classical instability we just mentioned, it can be seen as a collateral effect of Hawking radiation. It is the instability associated to the presence of an inner-outer horizon pair in stationary geometries, when characterized by differing surface gravities~\cite{Hollands:2019whz,McMaken:2023tft,McMaken:2023uue,McMaken:2024tpc,Balbinot:2023vcm}. 

\clearpage
Basically, the regularity of the vacuum at the outer horizon, normally imposed in determining the Hawking flux, is incompatible with regularity at an inner horizon with a different surface gravity. It can be shown (e.g.~in the Reissner--Nordstr\"om black hole~\cite{Balbinot:2023vcm}) that the instability  is associated to a diverging flux of the form $\hbox{(flux)} \propto (\kappa_{\rm inner}^2-\kappa_{\rm outer}^2)$. 
So, it will generically be present also for extremal inner horizons~\cite{Carballo-Rubio:2022kad,Franzin:2022wai} (which do not have exponentially fast classical mass inflation), whenever there is a nonzero surface gravity at the outer horizon~\cite{McMaken:2023uue}. 

Preliminary evidence points towards an associated instability with short timescales \cite{Barcelo:2020mjw,Barcelo:2022gii}
so that, at least for black holes far from extremality, such instability could be again at least as fast as the light crossing time (albeit the induced evolution might be much slower once the difference of the two surface gravities tends to zero).

With this picture in mind, it is clear that a fully time independent endpoint would have to be free of all of the aforementioned instabilities.
Herein we then argue for the following endpoint scenario: while we cannot exclude that there could be slowly evolving (meta-stable) near-extremal black holes or completely horizon-less objects, (which could be either naked singularities or gravastar-like objects~\cite{Mazur:2001,Mazur:2004,Mazur:2015,Visser:gravastar, Cattoen:2005,Holdom:2016}) only objects with one or more extremal horizons (and zero non-extremal horizons) could be the true (asymptotic) endpoints of gravitational collapse.

The structure of such objects would be very tightly constrained: purely on kinematical grounds the necessity for \emph{all} horizons to be extremal puts a very tight restriction on the spacetime geometry. Herein we shall describe a very general class of these fully-extremal spacetimes  for the static, non-rotating case, as well as for the the rotating one.  We also provide the line elements describing the associated horizon-less, ultra-compact, objects that might arise due to the classical light ring instability.

A final comment is due to the fact that the aforementioned instabilities are generically present also for black hole with regular cores (a possibility which indeed we shall consider in what follows) as long as they fall within the class endowed with inner horizons. A more extreme regularization scheme changing the topology of spacelike sections inside the black hole is possible (the so called black-bounce solutions~\cite{Simpson:2018tsi,Mazza:2021rgq,Franzin:2021vnj}) for which an outer horizon does not need to have a inner horizon counterpart as it entails in its interior a space-like wormhole throat~\cite{Carballo-Rubio:2019,Carballo-Rubio:2019fnb}. We shall see that this class of solutions can be recovered in our general scheme for fully extremal geometries.

\clearpage
\section{Spacetime geometry of fully extremal objects}
Let us, subject to suitable symmetry restrictions, seek to provide a reasonably general classification of fully extremal objects.

\subsection{Static case}
Perhaps the simplest example of a static spherically symmetric object with a single extremal horizon is the extremal Reissner--Nordstr\"om spacetime described by the line element
\begin{equation}
\label{E:extremal0}
\d s^2 = -  \left(1-{r_H\over r}\right)^2 \d t^2 + 
{1\over \left(1-{r_H\over r}\right)^2} \d r^2 + r^2 \d\Omega^2.
\end{equation}
Let us now seek to generalize this model spacetime in various ways.

First consider the line-element:
\begin{equation}
\label{E:extremal1}
\d s^2 = - \exp[-2\Phi(r)-2\Psi(r)] \left(1-{r_H\over r}\right)^2 \d t^2 + 
{\exp[2\Psi(r)] \over \left(1-{r_H\over r}\right)^2} \d r^2 + r^2 \d\Omega^2.
\end{equation}
Apart from the assumption of finiteness, the two functions $\Phi(r)$ and $\Psi(r)$ are allowed to be arbitrary. (And so the functions $\exp[\Phi(r)]$ and $\exp[\Psi(r)]$ possess neither zeros nor poles.) 

There is in this particular case a single extremal horizon at $r_H$ and this spacetime can be viewed as a distortion of extremal Reissner--Nordstr\"om, the distortion being encoded in  the two everywhere-finite functions $\Phi(r)$ and $\Psi(r)$. 

There is still generically a singularity at $r=0$, so to permit a regular core~\cite{Carballo-Rubio:2018,Carballo-Rubio:2019,Carballo-Rubio:2019fnb,Simpson:2019,Berry:2020} one {would need to further modify the geometry at small $r$. For example}
\begin{equation}
\label{E:extremal2}
\d s^2 = - \exp[-2\Phi(r)-2\Psi(r)] \left[ (r-r_H)^2\over r^2+r_0^2\right] 
\d t^2 + 
\exp[2\Psi(r)] \left[r^2+r_0^2\over (r-r_H)^2\right] \d r^2 + r^2 \d\Omega^2.
\end{equation}
The extra parameter $r_0$, when non-zero, allows for a regular core at $r=0$, while if $r_0\to 0$ one regains a generic singular core at $r=0$.
Note the Misner--Sharp quasilocal mass is
\begin{equation}
\label{misner-sharp-1}
m(r) = {r\over2} \left\{ 1 - \exp[-2\Psi(r)] \left[ (r-r_H)^2\over r^2+r_0^2\right] \right\},
\end{equation}
and that as long as $r_0\neq 0$ one has $\lim_{r\to0} m(r) = 0$.
 
This geometry can still be somewhat generalized. Consider now the line-element:
\begin{eqnarray}
\label{E:extremal3}
\d s^2 &=& - \exp[-2\Phi(r)-2\Psi(r)] \left[ (r-r_H)^2\over r^2+r_0^2\right]  \d t^2 
\nonumber\\
&& \qquad\qquad\;\; + 
\exp[2\Psi(r)] \left[r^2+r_0^2\over (r-r_H)^2\right] \d r^2 + 
\Xi(r)^2 \;\d\Omega^2.
\end{eqnarray}
Note the Misner--Sharp quasilocal mass is now
\begin{equation}
\label{misner-sharp-2}
m(r) = {\Xi(r)\over2} \left\{ 1 - [\Xi'(r)]^2 \exp[-2\Psi(r)] \left[ (r-r_H)^2\over r^2+r_0^2\right] \right\}.
\end{equation}

\enlargethispage{40pt}
The presence of the additional free function $\Xi(r)$ now allows for the possibility of wormhole throats~\cite{Morris:1988a,Morris:1988b, Visser:1989a, Visser:1989b, Simpson:2018, Franzin:2021}. Typically one would keep $\Xi(r)>0$, except possibly at $r=0$.

A local minimum of $\Xi(r)$ {located at $r>r_H$ corresponds to an ordinary traversable wormhole throat, whereas a  local minimum of $\Xi(r)$ located at $r_H$ corresponds to a wormhole with a null throat. A minimum at $r<r_H$ would correspond to a ``black bounce", i.e. to a spacelike wormhole throat inside the extremal horizon. }

This entire class of geometries is by construction stable against both Hawking radiation and mass inflation, {and the semiclassical instability at the inner horizon.
So, as anticipated, these geometries provides suitable candidates for the eventual endpoint of black hole evolution.}

A slightly more general model is to take $n\in\mathbb{Z}^+$ with $n\geq 2$ to be an otherwise arbitrary  positive integer,  and consider
\begin{eqnarray}
\label{E:extremal4}
\d s^2 &=& - \exp[-2\Phi(r)-2\Psi(r)] \left[ (r-r_H)^n\over r^n+r_0^n\right]  \d t^2 
\nonumber\\
&& \qquad\qquad\;\;+ 
\exp[2\Psi(r)] \left[r^n+r_0^n\over (r-r_H)^n\right] \d r^2 + \Xi(r)^2 \d\Omega^2.
\end{eqnarray}

One is now dealing with a higher-order $n$-fold degeneracy ($n\geq2$) at the single extremal horizon, and the geometry is again stable against both Hawking radiation and mass inflation.

Note that it is possible to consider a superfically  more general model by replacing $r^{n}+r_0^{n}$ with a more general polynomial of order $n$, a polynomial that has no real zeros in the region of physical interest. 
However, any such a polynomial could be absorbed into a redefinition of $\Psi(x)$, so this ``generalization'' would not actually add anything to the discussion of this essay.

\clearpage
A  more interesting and considerably more general model is to take the $n_i$ to be a set of positive integers ($n_i\in\mathbb{Z}^+$, with $n_i \geq 2$,  $i\in\{1,2,\dots, N\}$), and $r_{H_i}\in \mathbb{R}^+$,  and set
\begin{eqnarray}
\label{E:static}
\d s^2 &=& - \exp[-2\Phi(r)-2\Psi(r)] \;
\prod_{i=1}^N\left[ (r-r_{H_i})^{n_i}\over r^{n_i}+r_0^{n_i}\right] \d t^2
\nonumber\\
&&\qquad\qquad\;\;\;+ 
\exp[2\Psi(r)] 
\prod_{i=1}^N\left[r^{n_i}+r_0^{n_i}\over (r-r_{H_i})^{n_i}\right] \d r^2 + \Xi(r)^2 \d\Omega^2.
\end{eqnarray}
One is now dealing with multiple extremal horizons,  $N$ of them, located at $r_{H_i}$, possibly with higher-order $n_i$-fold degeneracies, allowing for the possibility of both wormhole throats and a regular core, 
and this entire class of spacetime geometries is again (by construction) stable against {all of the aforementioned instabilities}.

In all of these spherically symmetric static cases the line element is extremely tightly constrained, being characterized by a finite number of real and integer parameters, ($r_{H_i}$, $r_0$, $n_i$, and $N$), plus two finite but otherwise free functions, [$\Phi(r)$ and $\Psi(r)$], and an extra  function $\Xi(r)$ allowing for wormhole-like behaviour. The geometry  is presented in a form that is easily amenable to further investigation. 

Finally, note that the above (non-rotating spherically symmetric) discussion ended up focusing its attention on the specific rational polynomial function 
\begin{equation}
P(r) = \prod_{i=1}^N \left[ (r-r_{H_i})^{n_i}\over r^{n_i}+r_0^{n_i}\right]\,.
\end{equation}
When attempting to generalize our construction to the stationary (rotating) case, we shall soon see that the same polynomial will also play a crucial role.

\subsection{Stationary case}

When adding rotation, the line-element for the stationary case is considerably more subtle than that for the static case. 
The simplest extremal rotating object one might consider would be the (usual) extremal Kerr black hole, ($a\to m$), with the line element:
\begin{eqnarray} \label{extremal_kerr_metric}
\d s^2&=&  -\left(1-\frac{2mr}{r^2+m^2\cos^2\theta}\right)\; \d t^2
+\frac{r^2+m^2\cos^2\theta}{(r-m)^2}\; \d r^2\nonumber\\
&&+(r^2+m^2\cos^2\theta)\;\d\theta^2
-\;\frac{4 m^2r\sin^2\theta }{r^2+m^2\cos^2\theta}\;
\d t\d\phi\nonumber\\
&&+\left(r^2+m^2+{2m^3r\sin^2\theta\over{r^2+m^2\cos^2\theta}}\right)
\sin^2\theta\; \d\phi^2 .
\end{eqnarray}
But developing fully extremal models  a little more general than simple extremal Kerr would certainly be desirable. 

That is, one desires a model that is Kerr-like, rotating, but preserving as much as possible of the usual Kerr symmetries,  (including the Carter constant/Killing tensor and Klein--Gordon separability), while still  allowing for the presence of interesting and non-trivial extremal horizons.

\enlargethispage{40pt}
Starting from the general Kerr line element 
(that is, allowing $a\neq m$)~\cite{Kerr,Kerr-at-Texas,intro,book}
\begin{eqnarray} \label{kerr_metric}
\d s^2&=&  -\left(1-\frac{2mr}{r^2+a^2\cos^2\theta}\right)\; \d t^2
+\frac{r^2+a^2\cos^2\theta}{r^2-2mr+a^2}\; \d r^2\nonumber\\
&&+(r^2+a^2\cos^2\theta)\;\d\theta^2
-\;\frac{4 a mr\sin^2\theta }{r^2+a^2\cos^2\theta}\;\d t\d\phi\nonumber\\
&&+\left(r^2+a^2+{2mra^2\sin^2\theta\over{r^2+m^2\cos^2\theta}}\right)\sin^2\theta\; \d\phi^2,
\end{eqnarray}
a suitable 3-function extension of the Kerr spacetime is the line element~\cite{Baines:2023-3-function}:
\begin{eqnarray} \label{3f_kerr_metric}
\d s^2&=&  -\frac{\Delta(r)\exp[-2\Phi(r)]-a^2\sin^2\theta}{\Xi(r)^2+a^2\cos^2\theta}\; \d t^2
+\frac{\Xi(r)^2+a^2\cos^2\theta}{\Delta(r)}\; \d r^2\nonumber\\
&&+(\Xi(r)^2+a^2\cos^2\theta)\;\d\theta^2
-2\;\frac{a \sin^2\theta \;(\Xi(r)^2-\Delta(r)\exp[-2\Phi(r)]+a^2)}{\Xi(r)^2+a^2\cos^2\theta}\;\d t\d\phi\nonumber\\
&&+\frac{ \left((\Xi(r)^2+a^2)^2-\exp[-2\Phi(r)]\Delta(r)a^2\sin^2\theta\right)\sin^2\theta}{\Xi(r)^2+a^2\cos^2\theta}\; \d\phi^2 .
\end{eqnarray}

Here the functions $\Delta(r)$, $\Phi(r)$, and $\Xi(r)$ are at this stage essentially arbitrary. This line element is sufficiently general so as to allow one to describe a wide variety of Kerr-like spacetimes while still maintaining key symmetries, Killing tensors, and separability properties,  and (very importantly)  being able to easily calculate surface gravities~\cite{Baines:2023-3-function}.
\begin{itemize}
\item To maintain asymptotic flatness one must demand
\begin{equation}
\Delta(r)\sim r^2; \qquad \Phi(r)={\cal O}\left(1\right); \qquad \Xi(r)\sim r.
\end{equation}
\item
To recover standard Kerr one must demand
\begin{equation}
\Delta(r)\to r^2-2mr +a^2; \qquad \Phi(r)=0; \qquad \Xi(r) = r.
\end{equation}
\item
To recover standard extremal Kerr one must set $a \to m$ and demand
\begin{equation}
\Delta(r)\to (r-m)^2; \qquad \Phi(r)=0; \qquad \Xi(r) = r.
\end{equation}

\item For future comparison with the static case, the zero rotation ($a\to0$) limit is 
\begin{equation}
\d s^2 \to 
\exp[-2\Phi(r))]\;{\frac{{\Delta(r)}}{\Xi(r)^2}}\;\d t^2 + {\Xi(r)^2\over\Delta(r)} \; \d r^2 
+ \Xi(r)^2 [\d\theta^2 +\sin^2\theta \; \d\phi^2]. 
\end{equation}

\item 
For other Kerr-like spacetimes, see for instance references~\cite{Simpson:2021zfl,Simpson:2021dyo,Baines:2022xnh}.

\end{itemize}
In general, for this line element (\ref{3f_kerr_metric}) the Killing horizons are located at the roots $r_{H_i}$ of $\Delta(r)=0$, and these Killing horizons have angular velocities and surface gravities that (after a brief but slightly messy calculation) can be seen to be given by~\cite{Baines:2023-3-function}:
\begin{equation}
\Omega_{{H_i}}=\frac{a}{\Xi(r_{H_i})^2+a^2} ; \qquad 
\kappa_{H_i}=\frac{\exp[-\Phi(r_{H_i})]\;\Delta'(r_{H_i})}{2(\Xi(r_{H_i})^2+a^2)} .
\end{equation}

To now enforce extremality at all of the horizons one simply needs to enforce the condition $\Delta'(r_{H_i})=0$ at all of the horizons, 
which is tantamount to setting (with the $n_i\in\mathbb{Z}^+$ and $n_i\geq 2$)
\begin{equation}
\Delta(r) = \exp[-2\Psi(r)] \;  \Xi(r)^2\; \prod_{i=1}^N \left(1-{r_{H_i}\over r}\right)^{n_i}.
\end{equation}
Note we have not yet allowed for the possibility of a regular core at $r=0$.
The only constraint on $\Psi(r)$ is that is is finite, so that $\exp[\Psi(r)]$ has neither zeros nor poles.
Similarly one requires $\Xi(r)> 0$, except possibly at the ``center'' (if any) of the spacetime.
To allow for a regular core at $r=0$ one could introduce the extra parameter $r_0$ and set
\begin{equation}
\Delta(r) = \exp[-2\Psi(r)] \;  \Xi(r)^2\; \prod_{i=1}^N 
\left[ (r-r_{H_i})^{n_i}\over r^{n_i}+r_0^{n_i}\right].
\end{equation}

We again see (as for the static case) the prominent occurrence of the specific rational polynomial function 
\begin{equation}
P(r) = \prod_{i=1}^N \left[ (r-r_{H_i})^{n_i}\over r^{n_i}+r_0^{n_i}\right].
\end{equation}

Using $P(r)$ we can now write the 3-function fully extremal line element as 
\begin{eqnarray} 
\label{3f_kerr_metric_extremal}
\d s^2&=&  -\frac{ \exp[-2\Phi(r)-2\Psi(r)] \;  \Xi(r)^2\; P(r)-a^2\sin^2\theta}{\Xi(r)^2+a^2\cos^2\theta}\; \d t^2
\nonumber\\
&&
+\frac{ \exp[2\Psi(r)] \;(\Xi(r)^2+a^2\cos^2\theta)}{  \Xi(r)^2\; P(r)}\; \d r^2
+(\Xi(r)^2+a^2\cos^2\theta)\;\d\theta^2
\nonumber\\
&&
-2\;\frac{a \sin^2\theta \;(\Xi(r)^2- \exp[-2\Phi(r)-2\Psi(r)] \;  \Xi(r)^2\; P(r)+a^2)}{\Xi(r)^2+a^2\cos^2\theta}\;\d t\d\phi\nonumber\\
&&+\frac{ \left((\Xi(r)^2+a^2)^2-\exp[-2\Phi(r)-2\Psi(r)] \;  \Xi(r)^2\; P(r)a^2\sin^2\theta\right)\sin^2\theta}{\Xi(r)^2+a^2\cos^2\theta}\; \d\phi^2 .\qquad\quad
\end{eqnarray}

Because every horizon is now extremal, this spacetime is now stable against both Hawking radiation and mass inflation, while preserving the symmetries (including Carter constant, Killing tensor, and Klein--Gordon separability) of the Kerr spacetime it is modelled on. 
(Though this geometry does not generically preserve Dirac separability or the existence of a Killing--Yano tensor~\cite{Baines:2023-3-function}.) 
If we set $a\to0$ then the line element of equation (\ref{3f_kerr_metric_extremal}) is easily seen to reduce to that of the static case of equation (\ref{E:static}).
We emphasise that for this class of spacetimes the notion of extremity (zero surface gravity)
 is in general de-linked from the value of the spin parameter $a=J/m$. 
 
Overall, we now have a quite general but reasonably precise model for a rotating fully extremal black hole. 
The model is characterized by a finite number of real and integer parameters, ($a$, $r_{H_i}$, $r_0$, $n_i$, and $N$), plus two finite but otherwise free functions, [$\Phi(r)$ and $\Psi(r)$], and the free function $\Xi(r)$  controlling the possible presence of wormhole throats. The model is presented in a form easily amenable to further investigation. 

\section{Horizonless objects}
With the discussion for fully extremal spacetimes now completely under control, the situation for horizonless objects is straightforward, merely replace $P(r)\to 1$ (equivalently, set $n_i \to 0$) in order to eliminate all horizons. 
In the spherically-symmetric static case one finds
\begin{equation}
\label{E:horizonless-static}
\d s^2 = - \exp[-2\Phi(r)-2\Psi(r)]  \d t^2 + 
\exp[2\Psi(r)]  \d r^2 + \Xi(r)^2 \d\Omega^2,
\end{equation}
while in the Kerr-like stationary case
\begin{eqnarray} \label{E:horizonless-stationary}
\d s^2&=&  -\frac{ \exp[-2\Phi(r)-2\Psi(r)] \;  \Xi(r)^2-a^2\sin^2\theta}{\Xi(r)^2+a^2\cos^2\theta}\; \d t^2
\nonumber\\
&&
+\frac{ \exp[2\Psi(r)] \;(\Xi(r)^2+a^2\cos^2\theta)}{  \Xi(r)^2}\; \d r^2
+(\Xi(r)^2+a^2\cos^2\theta)\;\d\theta^2
\nonumber\\
&&
-2\;\frac{a \sin^2\theta \;(\Xi(r)^2- \exp[-2\Phi(r)-2\Psi(r)] \;  \Xi(r)^2+a^2)}{\Xi(r)^2+a^2\cos^2\theta}\;\d t\d\phi\nonumber\\
&&+\frac{ \left((\Xi(r)^2+a^2)^2-\exp[-2\Phi(r)-2\Psi(r)] \;  \Xi(r)^2\;a^2\sin^2\theta\right)\sin^2\theta}{\Xi(r)^2+a^2\cos^2\theta}\; \d\phi^2 .\qquad\quad
\end{eqnarray}
In view of the fact that both $\Phi(r)$ and $\Psi(r)$ were assumed finite, while $\Xi(r)>0$ except possibly at the center $r=0$, these geometries will certainly be horizon-free (while still preserving Klein--Gordon separability and the existence of a Killing tensor), {albeit possibly sporting a wormhole throat depending on the behaviour of the free function $\Xi(r)$.}

Taken in isolation, this \emph{form} for the line element (while it forbids horizons) will not yet distinguish nakedly singular spacetime geometries from gravastar-like everywhere nonsingular objects --- making such a distinction requires further explicit choices for the functions $\{\Phi(r),\Psi(r),\Xi(r)\}$, and further explicit computations of the curvature tensors.
{Additionally, horizonless objects can be subdivided into ultra-compact (possessing two or more light rings~\cite{Cunha:2017qtt,DiFilippo:2024}) and ``normal'' (without light rings and hence, for example, not exhibiting the shadows as seen by the EHT experiment nor the gravitational wave emission as seen by the LIGO-Virgo experiment). 

As already said, ultra-compact horizonless objects are expected to be generically unstable due to the inner light-ring instability. While we are lacking a general intuition about the back-reaction in this case, it is clear that an induced evolution towards a light-ring free configuration would not lead to an interesting class of black hole mimickers. Hence, one would generically demand that proposals for horizon-less black hole mimicker, will have to show an instability tending towards the formation of a horizon.}

\section{Discussion}

In this essay we have argued that the emergent scenario from the recently understood conundrum of instabilities affecting black holes with inner horizons, as well as their ultra compact counterparts, strongly suggest that fully-extremal black hole geometries could be the asymptotic, stable, end point of gravitational collapse. We then developed  various explicit line elements so to provide a solid, physically well-motivated, and mathematically tractable framework within which to more deeply analyze such evolution.

It turns out that the extremality conditions place quite severe kinematic restrictions on the spacetime geometry, and in this GRF Essay we have explicitly determined physically appropriate and mathematically tractable line-elements in both the static and stationary cases. The only alternative to all horizons being extremal is for there to be no horizons --- either in the form of naked singularities or gravastar-like compact objects. 
However, it is well known that such configurations if endowed with an outer light ring --- as necessary to fit the current VLBI experimental evidence for shadows --- will also be endowed with an inner (stable) light ring, which will hence lead to an instability.

Given this theoretical scenario, one might then wonder why most astrophysical black holes are not maximally rotating. Most importantly our family of extremal spacetimes~(\ref{3f_kerr_metric_extremal})  allows for zero surface gravity black holes  which are not maximally rotating. More subtly extremality might be achieved only asymptotically, given that astrophysical black holes are not in vacuum. We do not know if and how  such ``dirty black holes" can be subject to the aforementioned classical and semiclassical instabilities. Finally, even assuming that the instabilities operate in the same way for such ``dirty'' black holes, there is an issue of timescales. 
If the object tending to the fully extremal configuration is horizon-less, it will be evolving due to the classical light ring instability, for which no universal timescale is known. As we said, this could be astrophysically long. Similarly, if the meta-stable object is a black hole, it is unclear how efficiently it will evolve once it enters the  near extremal regime or it interacts with the surrounding matter.

In addition to all of the above, there is another relevant issue to be taken into account: the so called Aretakis instability associated to extremal horizons~\cite{Aretakis:2012ei,Aretakis:2011ha,Aretakis:2011hc,Murata:2013,Lucietti:2012xr}. This consists in a generic instability of the extremal horizon under scalar, electromagnetic or gravitational perturbations (something which might be even be prone to observational test in the future~\cite{Aretakis:2023ast}). 
Noticeably, it was numerically shown~\cite{Murata:2013} that, at least for an extremal Reissner-Nordstr\"om black hole, the nonlinear evolution of this instability generically makes the solution non-extremal. In some (fine-tuned) cases one can find a time-dependent extremal black hole, however, a freely falling observer will experience in this case arbitrarily large gradients in the field describing the initial perturbation at horizon-crossing, so that the horizon appears to these observers as effectively singular~\cite{Marolf:2010nd}.

One might take such preliminary results to reinforce the idea, derived from the third law of black hole mechanics, that fully-extremal configurations will be reached only asymptotically on very long timescales, as a result of the opposite competition of the aforementioned instabilities and possibly of astrophysical phenomena like accretion which might temporarily push the black hole away from extremality. Alternatively, they might be taken as a support for the more radical black-bounce, non-extremal solutions~\cite{Simpson:2018tsi,Mazza:2021rgq,Franzin:2021vnj} without inner horizons, as they appear (for now) much less prone to such instabilities.

In conclusion, it might well be that fully extremal configurations are the graveyard of gravitational collapse, but if they are so, it might still take a very long time for a black hole to die.

\bigskip
\bigskip
\hrule\hrule\hrule

\section*{Acknowledgements}
FDF acknowledges financial support by the Primus grant PRIMUS/23/SCI/005 from Charles University, and by the GAČR 23-07457S grant of the Czech Science Foundation. 

\bigskip
\hrule\hrule\hrule
\addtocontents{toc}{\bigskip\hrule}

\setcounter{secnumdepth}{0}
\section[\hspace{14pt}  References]{}
%

\end{document}